**Title:** Compressed Sensing with Signal Averaging for Improved Sensitivity and Motion Artifact Reduction in Fluorine-19 MRI

**Authors**: Emeline Darçot, PhD[1], Jérôme Yerly, PhD[1,2], Tom Hilbert, PhD[1,3,4], Roberto Colotti, PhD[1], Elena Najdenovska, PhD[1,2], Tobias Kober, PhD[1,3,4], Matthias Stuber, PhD[1,2], Ruud B. van Heeswijk, PhD[1,2]

[1]Department of Radiology, University Hospital (CHUV) and University of Lausanne (UNIL), Lausanne, Switzerland;

[2]Center for Biomedical Imaging (CIBM), Lausanne and Geneva, Switzerland;

[3]Advanced Clinical Imaging Technology (HC CMEA SUI DI PI), Siemens Healthcare AG, Lausanne, Switzerland;

[4]Signal Processing Laboratory 5 (LTS5), École Polytechnique Fédérale de Lausanne, Lausanne, Switzerland;

**Corresponding Author** (complete contact information):

Ruud B. van Heeswijk, PhD

Center for Biomedical Imaging (CIBM)

Centre Hospitalier Universitaire Vaudois (CHUV)

Rue du Bugnon 46, BH08.084

1011 Lausanne, Switzerland

Tel. +41-21-3147535

Ruud.mri@gmail.com

**Word count**: 7 386 words

**Running Head:** Compressed Sensing with Signal Averaging for Improved Fluorine-19 MRI

**Funding information:** This work was supported by grants from the Swiss Heart Foundation, the Swiss Multiple Sclerosis Society and the Swiss National Science Foundation (PZ00P3-154719 and 32003B_182615) to RBvH. Non-monetary research support was provided by Siemens Healthineers to MS.

**List of abbreviations:**

ApoE$^{-/-}$, apolipoprotein E-deficient;

BW, bandwidth;

CS, compressed sensing;

DSC, Dice similarity coefficient;

ETL, echo train length;

FSkC, fully sampled k-space center;

FWHM, full width at half maximum;

GT, ground truth;

ID, identity operator;

MSSIM, mean structural similarity index;

NAx-AFx, number of signal averages x with acceleration factor x;

PFC, perfluorocarbon;

PFCE, perfluoro-15-crown-5-ether;

PFOB, perfluorooctyl bromide;

PFPE, perfluoropolyether;

PSF, point spread function;

RMSE, root mean square error;

ROI, region of interest;

$SD_{noise}$, standard deviation of the background signal;

SNR, signal-to-noise ratio;

SPR, side-lobe-to-peak ratio;

$S_{SD}PR$, ratio of the standard deviation of the side lobe magnitudes to the main peak magnitude;

TV, total variation.

# Abstract

Fluorine-19 ($^{19}$F) MRI of injected perfluorocarbon emulsions (PFCs) allows for the non-invasive quantification of inflammation and cell tracking, but suffers from a low signal-to-noise ratio and extended scan time. To address this limitation, we tested the hypotheses that a $^{19}$F MRI pulse sequence that combines a specific undersampling regime with signal averaging has both increased sensitivity and robustness against motion artifacts compared to a non-averaged fully-sampled pulse sequence, when both datasets are reconstructed with compressed sensing. As a proof of principle, numerical simulations and phantom experiments were performed on selected variable ranges to characterize the point spread function (PSF) of undersampling patterns and the vulnerability to noise of undersampling and reconstruction parameters with paired numbers of x signal averages and acceleration factor x (NAx-AFx). The numerical simulations demonstrated that a probability density function that uses 25% of the samples to fully sample the k-space central area allowed for an optimal balance between limited blurring and artifact incoherence. At all investigated noise levels, the Dice similarity coefficient (DSC) strongly depended on the regularization parameters and acceleration factor. In phantoms, motion robustness of an NA8-AF8 undersampling pattern versus NA1-AF1 was evaluated with simulated and real motions. Differences were assessed with the DSC, and were consistently higher for the NA8-AF8 compared to the NA1-AF1 strategy, for both simulated and real cyclic motions (P<0.001). Both strategies were validated in vivo in mice (n=2) injected with perfluoropolyether. Here, the images displayed a sharper delineation of the liver with the NA8-AF8 strategy than with the NA1-AF1 strategy. In conclusion, we validated the hypothesis that in $^{19}$F MRI the combination of undersampling and averaging improves both the sensitivity and the robustness against motion artifacts.

# Introduction

Fluorine-19 ([19]F) magnetic resonance imaging (MRI) of injected perfluorocarbon emulsions (PFCs) is increasingly used for inflammation imaging and cell tracking.[1,2] Since [19]F is not naturally abundant in the human body, the [19]F atoms of the perfluorocarbon (PFC) can be directly quantified from the detected MR signal. In addition, several PFCs have been demonstrated to be safe for human use, and have already been injected with large volumes as blood volume expanders.[3] Given that they are taken up by immune cells, PFCs are also ideal biomarkers for inflammation sites, and [19]F MRI thus allows for a relatively straightforward quantification of their concentration (when influences on the relaxation times and $B_0$/$B_1$ fields are known or minimized).

However, since its MR signal only comes from the relatively low concentration of injected PFCs, [19]F MRI suffers from a low signal-to-noise ratio (SNR) that usually requires signal averaging to obtain interpretable images, which results in extended scan times. Several techniques have been investigated in order to overcome this challenge. This includes building optimized RF coils,[4] designing new PFCs with a high [19]F load,[5] and optimizing pulse sequence parameters.[6] If the emulsion has a multi-resonance spectrum, the SNR can be maximized through UTE acquisition,[7] deconvolution,[8] or chemical shift encoding.[9] However, in the case of challenging [19]F MRI applications that involve very small [19]F signals such as the detection of inflammation in atherosclerotic plaque[10,11] or tracking small quantities of injected cells,[2] these optimizations alone may still not suffice.

Another possibility to address the SNR limitation is the use of compressed sensing (CS) with signal averaging, as already qualitatively explored by several groups. CS consists of the iterative reconstruction of undersampled data, beyond the limit of the Nyquist-Shannon sampling theorem,[12] which must be sparse in a domain and must generate incoherent aliasing interferences in that sparse domain.[13] CS is commonly used to accelerate an acquisition. Given that the detected [19]F signal only comes from injected PFC, [19]F images tend to be sparse in the image domain directly, which makes them suitable for the application of CS. The combination of a [19]F acquisition with signal averaging and CS might provide two other major advantages: 1) an improvement of the sensitivity, i.e. the ability to accurately and precisely recover the small signals of low PFC concentrations, and 2) a gain in robustness against motion artifacts due to the signal averaging[14,15]. The principle of the first advantage might appear counterintuitive, given that an N-fold undersampled dataset (i.e. acceleration factor = N) that is averaged N times results in no net sample gain compared to a non-averaged fully-sampled dataset. However, an undersampling pattern that fully samples the k-space center and gradually undersamples the k-space periphery (Figure 1) will benefit from the property that most of the signal intensity is stored in the k-space center. Therefore, this scenario could provide an increased sensitivity and accuracy in the reconstructed image, when compared to a non-averaged fully sampled acquisition with an equivalent acquisition time. The second advantage, a reduced sensitivity to motion artifacts when averaging multiple acquisitions of the same k-space, may be well known for regular acquisition and reconstruction, but it is currently unclear whether and to what degree this benefit is maintained once the acquisition is semi-randomly undersampled.

Recently, several studies investigated applications of the combination of $^{19}$F MRI and CS, such as $^{19}$F catheter imaging[16] or cell tracking.[17] Zhong et al.[17] mainly investigated the gain in acquisition time enabled by CS, among others describing the beneficial denoising effect of CS when applied to a fully sampled dataset. Liang et al.[18] furthermore explored the efficiency of several CS algorithms at low SNR and concluded that the CS algorithm developed by Lustig et al.[13] remained the most efficient in terms of preserving the feature of the signal of interest. Several groups also explored the combination of CS and averaging, with different goals. Qualitative improvements in sensitivity[19] as well as the possibility to flexibly adapt the scan duration[20] were explored for $^{19}$F chemical shift imaging, while for high-resolution $^1$H imaging at low SNR it was shown to improve the spatial resolution[21,22]. Motion correction of $^{19}$F MRI was investigated by Keupp et al.,[23] who demonstrated the feasibility of motion correction by simultaneous $^1$H/$^{19}$F MR acquisition at 3T. They produced $^1$H and $^{19}$F motion-corrected images by applying motion tracking on sub-sampled $^1$H images. However, this method can only be applied to simultaneously acquired dual-nuclei acquisitions, which require highly specialized hardware.

Therefore, the hypotheses that the combination of signal averaging and undersampling results in improved sensitivity per acquisition time and in improved motion robustness compared to a non-averaged fully sampled acquisition, has to our best knowledge not yet been fully investigated, especially considering the complexities of image quality assessment when CS is applied.

The goal of this study was thus to investigate two hypotheses relating to the combination of CS and signal averaging with $^{19}$F MRI: 1) the increase in signal detection sensitivity per acquisition time, and 2) the improved robustness of the detected signal against motion artifacts, both compared to a non-averaged fully sampled dataset that is also reconstructed with CS. Since this is a proof-of-principle study, we do not propose to cover the entire range of possible acquisition-reconstruction strategies, nor to determine one generalized optimal strategy. Instead, we intend to cover and, to some extent, to optimize a defined range of variables in order to establish a strategy that allows a sufficient and fair evaluation of the two hypotheses. To this end, the first part of the study consisted of numerical simulations of several undersampling patterns to select a pattern with an optimal balance between the acceleration factor, the variable density of the undersampling, and the image fidelity. Then, in a phantom study, the influences of noise and of several motion patterns on the different investigated strategies were examined and quantified. Finally, a small in vivo animal study was performed to validate the in vitro findings.

## Methods

All imaging was performed on a 3T clinical MR scanner (MAGNETOM Prisma, Siemens Healthcare, Erlangen, Germany) with a 35-mm-diameter volume RF coil that is tunable to both the $^{19}$F and $^{1}$H resonances (Rapid Biomedical, Rimpar, Germany), and that was used for excitation and signal detection. An emulsion of the PFC polymer perfluoropolyether (PFPE, sold as VS-1000H by Celsense Inc, Pittsburgh, Pennsylvania, USA) was used for all experiments, since it has already been approved for clinical trials. According to the manufacturer, the emulsion had a $^{19}$F concentration of 4.20M (and thus a PFC concentration of 0.09M), a droplet size of ~180nm and a polydispersity of ~0.01.

All acquisitions were performed with an optimized isotropic 3D turbo spin echo (TSE) pulse sequence,[6] with field of view 32×32mm², slab thickness 32mm, slice oversampling 12.5%, voxel size 0.5×0.5×0.5mm³, echo train length 10, repetition time/echo time (TR/TE) 847/9.5ms, bandwidth 500Hz/px, acquisition time ~7 minutes and either with a predetermined undersampled trajectory or a fully sampled centric trajectory. The x direction is defined as the readout direction, while the y and z directions are defined as phase encoding directions. Undersampling will therefore always take place in the $k_y$-$k_z$ plane.

**Sampling pattern and trajectory design**

An algorithm was written in Matlab (The Mathworks, Natick, Massachusetts, USA) to design the undersampled k-space trajectories for maximal signal detection sensitivity and minimal eddy currents (due to large k-space jumps). First, a semi-random variable-density pattern with a fully sampled k-space center (with an adjustable radius) was generated.[24] The samples of this k-space were then divided into a number of equally populated concentric regions identical to the echo train length. The acquisition order of the echoes was then defined as follows: each echo from a given echo train was chosen from these regions, starting from the center to the periphery of the $k_y$-$k_z$ plane, with the first echo in the central region, given that it has the highest signal intensity. This trajectory is referred to as a center-out trajectory, as opposed to a traditional centric trajectory, where the progressive sampling of the echoes in an echo train will be performed from the k-space center to periphery in one phase encoding dimension only (Supporting Information Figure S1).

**Image reconstruction**

A previously published compressed sensing algorithm[13] was used for the reconstruction of both fully sampled and undersampled raw data with Matlab:

$$\arg\min_{m} \|\mathcal{F}_u m - y\|_2^2 + \lambda_\Psi \|\Psi m\|_1 + \lambda_{TV} \|\nabla m\|_1 + \lambda_{ID} \|m\|_1, \qquad (1)$$

where $m$ is the reconstructed image, $y$ is the acquired raw data, $\Psi$ is the wavelet operator (Debauchies-2 wavelet), $\nabla$ is the finite difference operator (the $\ell_1$-norm of $\nabla$ is also named total variation regularization (TV)), and the fourth term is the identity operator (ID). $\lambda_\Psi$, $\lambda_{TV}$ and $\lambda_{ID}$ are the matching regularization parameters, and $\mathcal{F}_u$ is the undersampled Fourier operator. In the case of fully sampled raw data, the fully sampled Fourier operator $F$ is used, and the algorithm behaves as a wavelet denoising filter:[25]

$$\arg\min_{m} \|Fm - y\|_2^2 + \lambda_\Psi \|\Psi m\|_1 + \lambda_{TV} \|\nabla m\|_1, \tag{2}$$

where both TV[26] and $\Psi$[27] are used as sparsifying transforms. The related regularization terms and the number of iterations were empirically established according to the reconstruction (with 100 and 32 iterations for CS and denoising reconstructions, respectively). These optimization problems were solved using the nonlinear conjugate gradient descent algorithm with backtracking line search.[13]

**Image quality assessment**

One of the challenges of CS and other iterative reconstruction techniques is to find a reliable and unbiased metric to quantify the quality and fidelity of the reconstructed image. The apparent noise in the reconstructed image is not true noise that comes from the data acquisition, but is transformed in the CS reconstruction process and highly depends on the chosen regularization terms. In this case, gold-standard measurement techniques such as the SNR that rely on the quantification of the standard deviation of the background noise of the image cannot provide a reliable measurement of the image quality. This has been addressed in several studies by using different image similarity metrics such as mean structural similarity index (MSSIM),[28] Dice similarity coefficient (DSC),[29] and root-mean-square-error (RMSE). These metrics, however, require a reference image, i.e. a ground truth (GT), to which the image that is evaluated is compared. Given that this is easily achievable in phantom experiments, DSC and RMSE were defined as image quality metrics for the phantom images. Conversely, the lack of ground truth made it inapplicable to the in vivo images of this study.

To perform the analysis with the DSC, a threshold was applied to both the ground truth and the test images to provide two binary masks of the objects of interest. The overlap between the corresponding pair of masks was then estimated with the DSC as follows:

$$\text{DSC(GT, Test)} = 2 \times \frac{|\text{GT} \cap \text{Test}|}{|\text{GT}| + |\text{Test}|}, \tag{3}$$

where $|\cdot|$ is the cardinality of the set, i.e. the number of voxels in the mask. A region of interest (ROI) was drawn in the object with the highest signal intensity in the image and the average signal intensity of this segmentation was calculated. The threshold used to calculate the binary masks was manually optimized in the ground truth image such that the created mask visually matched the phantom geometry and removed pixels from outside the phantom. This threshold was then applied to all images to create masks for the DSC calculation.

The RMSE was calculated as follows:

$$\text{RMSE} = \sqrt{\frac{\sum_{i=1}^{N}(Y_i - y_i)^2}{N}}, \tag{4}$$

where $Y_i$ was the $i^{th}$ pixel out of N of the GT image, and $y_i$, the $i^{th}$ pixel out of N of the Test image. Both the obtained DSC and RMSE were used to assess the differences between the tested strategies in terms of image quality.

### In silico study

Undersampling was performed in both phase-encoding directions, (the $k_y$-$k_z$ plane of k-space) on a simulated fully sampled unity 3D k-space, i.e. where all points were equal to 1. The undersampling patterns were optimized through the calibration of two parameters. The first parameter was the acceleration factor of the acquisition, which corresponds to the degree of undersampling of k-space and was set to be equal to the number of signal averages in this study. Besides, when it is combined with averaging and thus no longer results in a shorter acquisition time, it indicates the undersampling factor rather than the acceleration factor. Nine acceleration factors from 4 to 64 were investigated (acceleration factor = 4, 8, 16, 24, 32, …, 64), acceleration factors below 4 were omitted given results of previous studies[17,18]. The second parameter was the fully sampled k-space center (FSkC): the undersampling was performed with a variable-density function that fully sampled a k-space center area, i.e. the FSkC, and outside of which the periphery is progressively undersampled as a function of the distance to the center. This FSkC was calculated such that it contained either 1%, 12.5%, 25%, 37.5 or 50% of the total number of acquired k-space samples.

In order to choose an optimum combination of undersampling parameters that was neither too strongly affected by coherent undersampling artifacts nor by blurring effects, the point spread function (PSF) of the undersampling parameter combination was evaluated. The PSF of each of the 9×5=45 parameter combinations was calculated (by inverse Fourier transform, without iterative reconstruction) 100 times, with 100 different undersampling patterns to account for the randomness of the variable-density undersampling pattern simulations. Each PSF was evaluated by calculating the full width at half maximum (FWHM) in the central $k_y$-$k_z$ plane. For each combination of parameters, the average of the 100 FWHM values was used for comparison. As a measure of the incoherence generated by the undersampling patterns, the side-lobe-to-peak ratio (SPR)[13] and the ratio of the standard deviation of the side lobe magnitudes to the main peak magnitude ($S_{SD}PR$)[30,31] were calculated from the regional maxima in the central $k_y$-$k_z$ plane of each PSF. These measures inform on the amplitude and repetition of the potential coherent artifacts, respectively. Small values of both SPR and $S_{SD}PR$ indicate low coherence of the aliasing artifacts.

Given that the ordering of the acquired echoes was centric for the fully sampled acquisition, used for the retrospective undersampling, versus radially center-out for the prospectively undersampled acquisition, retrospective and prospective undersampling will affect image quality differently, especially when $T_2$ relaxation is taken into account. Therefore, to evaluate the effect of different $T_2$ decays on the blurriness of the reconstructed image, numerical simulations were performed with three different PFCs: perfluoropolyether emulsion (PFPE, $T_2$=155±12ms at 24°C),[6] perfluorooctyl bromide emulsion (PFOB $T_2$=283±20ms),[6] and perfluoro-15-crown-5-ether emulsion (PFCE, $T_2$=588±28ms). For all three PFCs, each echo was multiplied with a $T_2$ decay coefficient according to its echo number, the used trajectory, and the sequence timing. This effect was characterized for an undersampling pattern with acceleration factor 8 and FSkC 25%. The PSF of both simulated trajectories was calculated ten times, with a zero-filled reconstruction. The FWHM was then used to evaluate the blurring effect of the two different trajectories in the $k_y$-$k_z$ plane at the center of k-space. $T_2$ relaxation times of the PFC emulsions at 24°C were used for these simulations, since most of the quantitative experiments in this study were performed in phantoms at room temperature.

### In vitro study

A phantom was constructed with five 1-mL syringes of agar gel mixed with PFPE emulsion at different $^{19}F$ concentrations (1.05M, 0.52M, 0.26M, 0.13M and 0M). These syringes were embedded in a 50-mL tube filled with agar gel. Since the effect of agar on PFC properties is minimal and is the same for all tubes,[6] we assumed that there was no net effect on the quantification due to differences in $T_2$ relaxation times between the PFPE emulsion and the PFPE emulsion-agar mix.

### *Noise simulations*

In order to characterize the undersampling and averaging combination with regard to noise while their interaction with the scanner hardware was also included, a series of datasets with different noise levels was generated from a fully sampled 64 averages static acquisition dataset, which was also used as a ground truth. A noise sample was added to each point in k-space, and effect of this noise on each of the 32 imaging slices (which all contain the same tubes) was quantified and averaged. Three datasets were generated with an SNR of 17, 8 and 4 in the syringe with the highest $^{19}F$ concentration. Based on the results of the numerical simulations, reconstruction with several parameter combinations were simulated and tested with acceleration factor = 4, 8 and 16 (and corresponding averages) and all with FSkC of 25%. The three acceleration factors were retrospectively applied to the three noise level datasets. To each of these nine acceleration-SNR combinations, a series of reconstruction parameter combinations was applied: in total, nine combinations of the regularization parameters $\lambda_\Psi$, $\lambda_{TV}$ and $\lambda_{ID}$ were tested (0.001, 0.005, 0.01 for both $\lambda_\Psi$ and $\lambda_{TV}$; $\lambda_{ID}$ was set to 0.01). Each reconstruction was allowed 100 iterations.

A quantitative measure of the image quality was then obtained by calculating the DSC. The threshold was set to 3% of the average signal intensity of the brightest syringe in each image. The ground truth was defined as the original NA=64 acquisition denoised through a wavelet denoising filter with $\lambda_{TV}$=0.05 and $\lambda_\Psi$=0.05 to obtain a clean mask of the four tubes. The DSC was calculated to assess the differences between the tested combinations in terms of image quality.

To assess the fidelity of the reconstructed signal intensities at different noise levels, ROIs of 50 pixels were drawn for each syringe in the central ground truth slice and applied to all tested images. Linear fits of the signal intensities of the five phantom syringes as a function of their $^{19}F$ concentration were then made, and the coefficient of determination ($R^2$) was calculated.

The root-mean-square error (RMSE) was also calculated to compare the image quality between the different undersampling-averaging and reconstruction parameter combinations at the three different noise levels. Beforehand, images were normalized to the highest signal intensity in the image, which was calculated as the mean of the highest 1% intensities in order to avoid potential outliers.

*Motion simulations*

Both simulated and real motion were applied to a fully sampled non-averaged TSE acquisition with a centric trajectory (NA1-AF1) and a prospectively undersampled acquisition with eight short-term averages and a center-out k-space trajectory, acceleration factor 8,[17,20] and FSkC=25% (NA8-AF8). Since real motion is never perfectly consistent with simulated motion (due to non-linear 3D movement, not fully reproduced motions and frequency, etc.), the goal of the real motion experiments was to confirm the improved performance of the undersampled-average reconstruction rather than to quantitatively reproduce the simulations. The regularization parameters used for the CS reconstruction were $\lambda_{TV}=0.003$, $\lambda_{\Psi}=0.005$, and $\lambda_{ID}=0.07$, while $\lambda_{TV}=0.03$ and $\lambda_{\Psi}=0.08$ were used for the denoised reconstruction.

The simulated motion was added as a linear k-space phase shift to the raw data before the reconstruction. Three different motion patterns were used: an approximation of a sudden permanent whole-body movement (body motion, Figure 2a) at half of the acquisition time ($T_{acq}$), a sinusoidal motion with a period P=1200ms (sine motion, Figure 2b), and an asymmetric periodic motion that models a breathing regime, with a short inspiration (25% of motion) and a long constant end-expiration (breathing motion, Figure 2c), with a period P=1500ms (i.e. 40 breaths per minute – bpm), which is a typical value for anesthetized small animals. These motion types were expected to result both in blurring if the signal would be incoherently spread out, and in ghosting if coherent motion states were generated, either through synchronization between the motion and TR, or simply due to repeated acquisition during an often occurring motion state such as end-expiration. All motion patterns were applied in one phase-encoding dimension, and their amplitudes were normalized to 10%, 1% and 30% of the field of view, matching a displacement of 3.2mm, 0.32mm and 9.6mm, for body, sine, and breathing motion, respectively. A range of amplitudes (10 to 100% of the FOV) and frequencies (from 20 to 200 bpm) were investigated for the breathing motion. Motion simulations were also performed with the real motion frequencies and amplitudes described below.

The real motion was applied during scanning with a manual pump and a 3L inflatable water reservoir (Platypus, Seattle, Washington, USA) placed beneath the coil in order to mimic the three investigated motions: body motion, sine motion, and breathing motion. The amplitudes were 1.1cm, 0.5cm and 1cm, respectively (34%, 16% and 31% of the field of view, respectively). For the cyclic motions (sine and breathing motions), the frequencies were 11.6 periods per minute and 1.5 periods per minute (periods of 5.2s and 40s, respectively). The actual movement of the breathing motion occupied 30% of the period. These motion amplitudes were validated by a real-time balanced steady-state free precession (bSSFP) cine acquisition of the three movements. The motion was applied in the plane perpendicular to the length of the syringes (i.e. transversal plane) for a better visualization of the motion, in both phase-encoding directions that were undersampled. The strategies were the same as for the simulated motion equivalents.

The DSC was calculated to quantify the degree of motion compensation of the NA8-AF8 acquisition compared to the denoised NA1-AF1 acquisition for both the simulated and real motions. For both acquisition strategies, the corresponding static acquisition was defined as the ground truth. The mask threshold was set to 7% and 3.5% of the average signal intensity of the brightest syringe for NA1-AF1 and NA8-AF8, respectively.

The RMSE was also calculated to compare the image quality between NA1-AF1 denoised images and NA8-AF8 images for both the simulated and real motions. As for the noise simulation experiments, images were normalized beforehand to the highest signal intensity in the image, which was calculated as the mean of the highest 1% intensities in order to avoid potential outliers.

In order to assess the quantification accuracy and the sensitivity of the various reconstruction techniques, the dataset, which present an SNR of 15 in the tube with the highest signal intensity, was used to generate two additional datasets with an SNR of 8 and 4 in the tube with the highest signal intensity, as performed in the noise simulations. For both strategies that require iterative reconstruction, the same regularization parameters were kept for the reconstruction of the images with different SNR levels. For the signal intensity measurement of the five tubes, ROIs were drawn in the tubes in the ground truth image and used for all strategies. The signal intensity in each tube for all three strategies and all SNR levels was plotted against the known $^{19}$F concentration, linear fits were made, and the goodness of the fits ($R^2$) was calculated.

## In vivo study

To validate the in vitro results, two C57BL/6 apolipoprotein E-deficient (ApoE$^{-/-}$) mice, which are a model of hypercholesterolemia and atherosclerosis,[32] were scanned one day after an intraperitoneal injection of 300μL of PFPE. Permission from the local Animal Ethics Committee was obtained for all animal experiments performed in this study. The animals were anesthetized with 1.5-2% of isoflurane in 100% oxygen during the scan. Body temperature and respiration rate were monitored with a rectal probe and a respiration pillow that was placed below the chest of the mouse (SA Instruments, Stony Brook, New York, USA).

After $^1$H GRE acquisition for anatomic localization of the liver and spleen, $^{19}$F TSE acquisitions were performed with both NA1-AF1 and NA8-AF8. The regularization parameters used for the CS reconstruction were: $\lambda_{TV}$=0.005; $\lambda_\Psi$=0.001; $\lambda_{ID}$=0.04; and $\lambda_{TV}$=0.025; $\lambda_\Psi$=0.01 for the denoised reconstruction.

## Statistical analysis

All continuous variables are reported as average ± standard deviation. For the phantom analysis, a paired Student's t-test with a Bonferroni correction for multiple comparisons was used to account for significant differences in DSC between the different noise floors or acquisition strategies, with P<0.05 considered significant. The same was performed for the RMSE. A repeated-measures ANOVA was performed to assess the significant effect of the different parameters of the noise simulations on the DSC.

# Results

All Figures with phantom images show the y-z plane in which the undersampling was performed. The in vivo images, which were acquired in a different orientation, show the y-z plane as a sagittal view and x-y plane as a coronal view.

## Sampling patterns and acceleration

The FWHM of the PSF of a series of undersampling patterns increased with the acceleration factor and the FSkC (Figure 3a, Supporting Information Figure S2). For FSkC=50% for example, the FWHM increased from 2.08±0.01 to 8.67±0.33 pixels. Both measures of incoherence (SPR and $S_{SD}PR$) demonstrated a decrease in the incoherence of the undersampling artifacts with the increase of the acceleration factor (Figure 3b and c). In both graphs, the same behavior was observed: the 12.5% FSkC curve crossed the 1% FSkC curve at acceleration factor =32, indicating a higher coherence in aliasing artifacts at high acceleration factor for a FSkC=12.5% pattern than for a FSkC=1% pattern. However, the incoherence measurement increased with the FSkC (from SPR=0.104±0.008 to SPR=0.078±0.005 from 1% to 50%, for example at acceleration factor =8; Figure 3b and c). Based on these results, only the lowest three acceleration factors were kept for the remainder of the study. Given that a low FSkC results in limited blurring and that a high FSkC increases the artifact incoherence required for CS, we chose to use the mid-range FSkC of 25% for the in vitro and in vivo studies.

## $T_2$ relaxation

When $T_2$ relaxation during the acquisition was included in the simulation, the FWHM of the PSF increased slightly more for echoes acquired with the center-out trajectory than for the centric trajectory (Figure 3d). Both trajectories with simulated $T_2$ decay resulted in higher FWHMs than the equivalent undersampled k-space without simulated $T_2$ decay (Figure 3d). When the longer $T_2$ relaxation times of the perfluorocarbons PFOB and PFCE were used, the FWHM increase was even less pronounced.

## SNR and regularization parameters

The fidelity of the reconstructed image as assessed with the DSC in the phantoms significantly depended on the regularization parameters (P<0.001), as well as on both the SNR in the original dataset and the acceleration factor (P<0.001). At the highest SNR level (SNR=17), the DSC varied more as a function of the regularization parameters at acceleration factor 4 than at higher acceleration factor (Figure 4a). However, at SNR=8 the regularization parameters had a stronger influence on the DSC at acceleration factor 8 than at acceleration factor 4 or 16. Finally, at SNR=4 the regularization parameters had a stronger effect at acceleration factor 16. The DSC of the datasets with lower SNR were significantly different from the DSC of the SNR=17 dataset for all acceleration factor comparisons (P<0.001) except for SNR=8, acceleration factor 16 (P=0.07). These image quality assessments were confirmed by the RMSE calculations, where the RMSE of SNR=8 and 4 were significantly different from those of SNR=17 for all parameter combinations (P<0.04, Supporting Information Figure S3).

The coefficient of determination ($R^2$) increased together with the SNR (Figure 4b). For all the 27 acquisition-reconstruction strategies, the $R^2$ values at SNR=8 were not significantly different from the $R^2$ values at SNR=17, while it was significantly different compared to $R^2$ values at SNR=4 (P>0.5 and P<0.001, respectively). Averaged over the different reconstruction parameter combinations, $R^2$ was slightly higher at acceleration factor 16 than at acceleration factor 8 or 4, at all SNR levels (Figure 4b); all SNR and parameter combinations included, $R^2$ values were significantly different between all acceleration factors (P<0.01).

## Simulated and real motion

A clear reduction of the background signal was observed in the NA8-AF8 images compared to the denoised NA1-AF1 images for both cyclic motions (sine and breathing). However, no major difference could be observed between the NA8-AF8 and denoised NA1-AF1 images for both the simulated (Figure 5) and real (Figure 6) body motion. While the artifacts in simulated sine and breathing motion images were coherent, i.e. several ghosting syringes could be observed, in the real-motion images the artifacts mostly consisted of added background noise.

For cyclic motion, the DSC of the NA8-AF8 images was consistently higher than that of denoised NA1-AF1 images ($P<0.001$, Table 1), while it resulted in lower DSCs for body motion ($P<0.001$, Table 1). These image quality trends were confirmed by the RMSE values (Table 2).

The breathing motion amplitude had only minor impact on the resulting images and on the DSC for both strategies ($DSC_{NA1-AF1}$ range = [0.31 - 0.38]; $DSC_{NA8-AF8}$ range [0.75 - 0.82]). The frequency variation had a similar small visual effect on the images, with an interesting exception at 140 bpm, which is almost exactly double the frequency of the acquisition (i.e. the motion period is half the TR). At this specific frequency a large increase in ghosting was observed, which led to a much lower $DSC_{NA8-AF8}=0.36$ than the rest of the range ($DSC_{NA1-AF1}$ range = [0.26 - 0.36], $DSC_{NA8-AF8}$ = [0.77 - 0.84], Supporting Information Figures S4, S5, S6).

The motion simulations with real motion parameters resulted in images and DSCs of both the body and breathing motions that confirmed the findings of the simulated and real motion experiments. The sine motion resulted in images without recognizable signals for both NA1-AF1 and NA8-AF8 strategies (Supporting Information Figure S7).

Linear regressions resulted in $R^2$ values between 0.9339 for NA1-AF1 strategy at SNR=8 and 0.9964 for NA1-AF1 denoised at SNR=15 (Figure 7). At SNR=15, the main difference between the three concentrations versus signal plots appeared to be the signal intensity of the fifth point, which is supposed to be zero, but due to unsuppressed noise was non-zero for the standard reconstruction. While the iterative strategies (NA1-AF1 denoised and NA8-AF8) have a similar performance at SNR 15 (as measured in the image without iterative reconstruction), the NA8-AF8 strategy outperforms NA1-AF1 at the lower SNR=4, and thus improves the sensitivity of $^{19}F$ MRI (Figure 7).

### In vivo validation

In vivo, the PFC-loaded liver and spleen also identified in $^1H$ image (Figure 8a) were clearly visible in the denoised NA1-AF1 images and NA8-AF8 images with CS reconstruction (Figure 8e-g and h-j, respectively), while it was barely distinguishable from the noise in NA1-AF1 images without iterative reconstruction (Figure 8b-d). The delineation of the liver and spleen in the three visible slices of the NA8-AF8 strategy furthermore appeared sharper on visual inspection than in the NA1-AF1 images (Figure 8e-j). Qualitatively, the conspicuity of the local intensity variation in the liver and spleen was improved without losing detailed information.

## Discussion

Several undersampling patterns and acquisition-reconstruction strategies were tested and characterized. After comparison of these patterns and strategies, an optimized NA8-AF8 strategy was selected for the evaluation of our hypotheses. Both hypotheses were confirmed: the NA8-AF8 strategy demonstrated a better sensitivity and robustness against cyclic motion artifacts than a denoised fully sampled non-averaged strategy. A possible explanation for the improved performance of the undersampled-average acquisition is that the k-space points that are averaged have a better SNR, and thus stand out much more readily from the noise when (soft) thresholding is used in the iterative reconstruction algorithms. More signal will then be correctly represented in the final image in the undersampling-averaging case compared to the denoising reconstruction. Simultaneously, the (soft) thresholding will remove the noise in a similar manner in both reconstructions, resulting in a net improved undersampled-averaged reconstruction. This was recently also confirmed for low-SNR $^1$H imaging with large matrices and variable density averaging, where the number of averages depended on the proximity to the k-space center[21].

All in silico optimizations were performed directly with the PSF, and are thus a global reflection of the ensuing spatial resolution, although it should be noted that severe peripheral undersampling may result in local decreases in spatial resolution around small structures[33]. The PSF simulations demonstrated that the FWHM increased with the acceleration factor. This led us to keep only the lowest three acceleration factors (4, 8 and 16) for the in vitro study. Similarly, FSkC=25% was selected to balance limited blurring in the image with the high incoherence required for CS.

The small increase of the FWHM with the center-out trajectory when the $T_2$ decay was included in the PSF simulations occurred due to the temporally coherent distribution of the signal intensity in k-space: While low k-space frequencies were only sampled by early echoes of the echo train, high frequencies were sampled by late echoes. This created a low-pass filter effect on k-space as previously described by Tamir et al.[34] This effect was stronger when a lower $T_2$ value was used due to the increased difference in signal intensity between early and late echoes of one echo train. The optimized trajectory for increased signal detection thus comes at the cost of some blurring in the images. However, it should be noted that these high FWHM values were obtained from PSFs that were reconstructed without any iterative reconstruction and might be partially compensated by the CS reconstruction.

In the study on the effect of noise levels, both the DSC and the coefficient of determination $R^2$ consistently increased or stayed at the maximum when the degree of undersampling-averaging was increased, except at the highest SNR. Increasing the regularization parameters tended to result in a higher DSC. Nevertheless, this has to be balanced with the risk that an over-regularization might induce a smoothing and blurring of the image, which might then distort or conceal details of the image. This tradeoff between regularization and denoising needs to be carefully calibrated, especially when more complex geometrical structures are investigated, in order to avoid any loss of information in the detected $^{19}$F signal. Previous studies found that at low SNR, low degrees of CS acceleration give better results.[17,18] However, our results demonstrate that at an SNR of 4, the acceleration factor 4 datasets consistently had the lowest DSC. This might be explained by the fact that the DSC focuses on the geometry and not on the blurring of the image: a blurred image with reduced background noise will often result in a higher DSC than a noisy image with the original edge sharpness. Although the DSC is highly appropriate for the evaluation of the relatively simple phantom structure used in this study, it thus only assesses a part of the overall image quality, which prompted us to add the RMSE as an additional image quality metric in this study.

Overall, NA8-AF8 most consistently outperformed the other averaging-undersampling combinations in silico and in vitro experiments, and was chosen for the in vitro and in vivo motion experiments. This also agreed with the previous findings of Zhong et al.[17] The undersampling-averaging sampling strategy reduced ghosting artifacts from cyclic motions, since the motion is incoherently spread over the averaged samples, which smoothens and cancels out the different motion states of the phantom. This observation combined with the CS reconstruction most likely explains the difference observed between NA1-AF1 and NA8-AF8 cyclic motion DSC. For the non-cyclic body motion, this incoherent spreading does not occur, since there are two coherent motion states that cannot be compensated by averaging, and thus appear as two overlaid shifted and incompletely sampled images. Hence, a difference in DSC of the NA8-AF8 images is observed between cyclic and non-cyclic motion. The NA1-AF1 images furthermore confirm this, since without averaging, all three types of motion have similar DSCs.

The difference in artifacts between the simulated and real cyclic motion images can be at least partly explained by the different amplitudes and frequencies of the corresponding motions, which had to be used because of the physical constraints of the moving phantom experiments, as well as a small spread in the speeds, displacements, and durations of the human-driven and non-ideal real motion. However, it should be noted that it was not the intention to reproduce the exact motion of the simulations with the real motions, but only to demonstrate generally similar results. A further cause might be the intra-readout motion that occurred during the real motion acquisition, while we did not add any for the simulated motion. The frequency at which the object of interest moves relative to the acquisition also plays a role in the degree of motion artifact reduction, but remains independent of the acquisition technique. Therefore, in future studies, the acquisition parameters could potentially be adapted to the motion frequency of the subject (when known) while keeping them within a range that results in maximum signal strength. To this end, as a future step, the existence of a mathematical relationship between the motion frequency and sequence timing could be investigated. It should also be noted that all motion types will most likely be non-rigid in vivo, and will result in different displacements throughout the body. The resulting varying levels of blurring will lead to differently lowered local signals and thus to a spatially varying underestimation of the concentration, as already observed by Keupp et al[23]. As explored in that study, this might be partially corrected for by reconstructing sub-images, or in our case images of the single averages, and performing registration on these sub-images. However, this might perform better for rigid or one-time motion than for cyclic motion that would equally affect all sub-images.

The close regression curves of all three strategies used for motion simulations confirm that regularization does not affect the concentration quantification. The higher $R^2$ of the NA8-AF8 and NA1-AF1 denoised strategies indicate that they were superior to those of the regular reconstruction. However, this might be purely due to the lack of signal of the proposed reconstructions in the syringe without PFC.

Initial tests (data not shown to limit the number of reported optimization steps) showed that image quality improved when the wavelet regularization was added to the CS reconstruction. This occurred despite the sparsity of the $^{19}$F MR images directly in the image domain, most likely because the signal still took up non-negligible space in our in vitro and in vivo images, and because the wavelet domain allows for more efficient compression. Therefore, while it has not been used in previous $^{19}$F MR studies, we chose to include wavelet regularization in our CS algorithm. Kampf et al.[20] also investigated the use of non-convex $\ell_p$-norms (p<1) that are more efficient in noise-free datasets but induced more spike artifacts in noisy datasets. They recognized that in presence of low SNR, p=1 would still provide the best results. Considering the complexity of using a non-convex norm, we used the $\ell_1$–norm for the minimization.

One acquisition parameter that might benefit the averaged-undersampled method, but was not investigated in this study, is the way the averaging was performed. A short-term averaging mode was used during the acquisition: for a N-average acquisition, each k-space line was acquired N times before acquiring the next. With a long-term averaging mode, the entire k-space is acquired once before acquiring the next k-space. Using a long-term averaging mode to compensate for a one-movement motion, like our body motion, might still not fully compensate for it, but as the motion will be better distributed over all averaged k-space lines, this might result in a higher conspicuity of the object, even though this mode also depends on the motion period. Another aspect to investigate in further work is the use of a bSSFP pulse sequence. This study was performed with a TSE pulse sequence that was chosen for providing a high SNR. However, a bSSFP sequence can be used to obtain a higher ratio SNR/time efficiency compared to TSE and might be of interest. Similarly, Cartesian sampling was chosen for this study in order to ensure a high SNR efficiency per unit time instead of radial sampling, which might have provided a stronger robustness to motion artifacts. The undersampling-averaging sampling strategy was also briefly investigated in $^1$H carotid imaging[22] where this allows for a higher resolution. Interestingly, the abovementioned study by Schoormans et al.[21] explored CS with averaging that increased or decreased depending on the proximity to the k-space center, and demonstrated that increasing the averaging with the proximity to the k-space center further improved the image quality in their high-resolution $^1$H images. An additional method to combine to this one to improve the CS image quality is that of Kampf et al., who investigated two different post-processing resampling strategies to reduce the spike artifacts due to the undersampling and without the need of additional data acquisition.[35]

The main limitation of this study is the inherent incomplete exploration of the parameter space: we set out to illustrate that averaging and compressed sensing improve sensitivity and motion robustness, not to establish absolute optimal recipes for $^{19}$F MRI with CS and averaging. Furthermore, an optimal undersampling pattern and parameter set would only be useful for a single type of image. Indeed, Zijlstra et al. suggests that the optimal sampling density depends on the acquired image, and furthermore that using a suboptimal undersampling pattern would lead to a lower reconstruction quality.[36] Therefore, this study was designed as an exploratory study in which we analyzed and characterized several aspects of multiple acquisition-reconstruction strategies for the acquisition of prospective undersampled raw data on a moving subject with at least some degree of optimization. For instance, the three weights for each regularization term in the CS reconstruction were chosen to cover a large range of potential reconstructions, but could still be fine-tuned to improve the reconstruction. These ranges thus resulted in different parameters for the motion simulation reconstructions and the in vivo image reconstructions. Nevertheless, the finding of optimal regularization parameters and FSkC was beyond the scope of this study. Therefore, further investigation on the weight combinations for each reconstruction could lead to slightly different or even improved results. Furthermore, with the onset of machine learning[37] and the arrival of a new generation of optimization algorithms for CS in MRI (such as ADMM),[38] the idea of analytical optimization of the regularization parameters for each combination of image acquisition and reconstruction could be envisaged as a step toward a more informed use of CS.[39] A second limitation is the absence of a standard method to quantify the detection limit (i.e. the lowest cutoff concentration that generates an identifiable signal) of regularized images. Given the regularization of the background noise, unrealistically low detection limits would be obtained with standard techniques such as the Rose criterion,[40] which is why no cutoff values were calculated. Only rigid translational motion was investigated in the simulations and phantom studies, since this is what smaller structures such as inflamed tissues typically undergo. Finally, the design of our phantom tubes (4-5mm diameter) with homogeneous PFC distributions (required to have well-characterized references) did not enable us to investigate the sensitivity provided by our technique beyond the millimetric level nor the effect of inhomogeneously distributed signals.

In conclusion, , we validated the hypotheses that an N-fold undersampled acquisition with N averages improves both the sensitivity of the signal per unit time and the robustness against cyclic motion artifacts compared to a non-averaged fully sampled dataset when both were reconstructed with compressed sensing, in the context of a defined undersampling pattern, structure, and averaging range.

## Acknowledgements

We would like to thank Maxime Pellegrin, PhD for his assistance with the animal experiments, Jean-François Knebel, PhD and Jean Delacoste, PhD for their assistance with the statistical analysis, as well as Graham A. Wright, PhD for his valuable discussions on the study. This work was supported by grants from the Swiss Heart Foundation, the Swiss Multiple Sclerosis Society and the Swiss National Science Foundation (PZ00P3-154719 and 32003B_182615) to RBvH. Non-monetary research support was provided by Siemens Healthineers to MS.

# Tables

**Table 1. Dice similarity coefficients (DSCs) of the simulated and real phantom motions.** The DSCs were calculated between each image with induced motion and their corresponding static image, where both were reconstructed with the same reconstruction parameters. All DSCs were significantly different between the two sampling strategies (P<0.001).

| DSC [-] | Simulated motion | | Real motion | |
| --- | --- | --- | --- | --- |
| | NA1-AF1 | NA8-AF8 | NA1-AF1 | NA8-AF8 |
| Body motion | 0.37 | 0.26 | 0.35 | 0.18 |
| Sine motion | 0.42 | 0.86 | 0.26 | 0.48 |
| Breathing motion | 0.31 | 0.77 | 0.32 | 0.35 |

**Table 2. Root-mean-square error (RMSE) of the simulated and real phantom motions.** The RMSE was calculated between each image with induced motion and their corresponding static image, where both were reconstructed with the same reconstruction parameters. A lower value indicates higher image quality. The differences between NA1-AF1 and NA8-AF8 values agree with the DSC results.

| RMSE [-] | Simulated motion | | Real motion | |
| --- | --- | --- | --- | --- |
| | NA1-AF1 | NA8-AF8 | NA1-AF1 | NA8-AF8 |
| Body motion | 0.12 | 0.14 | 0.11 | 0.14 |
| Sine motion | 0.10 | 0.04 | 0.15 | 0.11 |
| Breathing motion | 0.08 | 0.04 | 0.08 | 0.05 |

# Figures

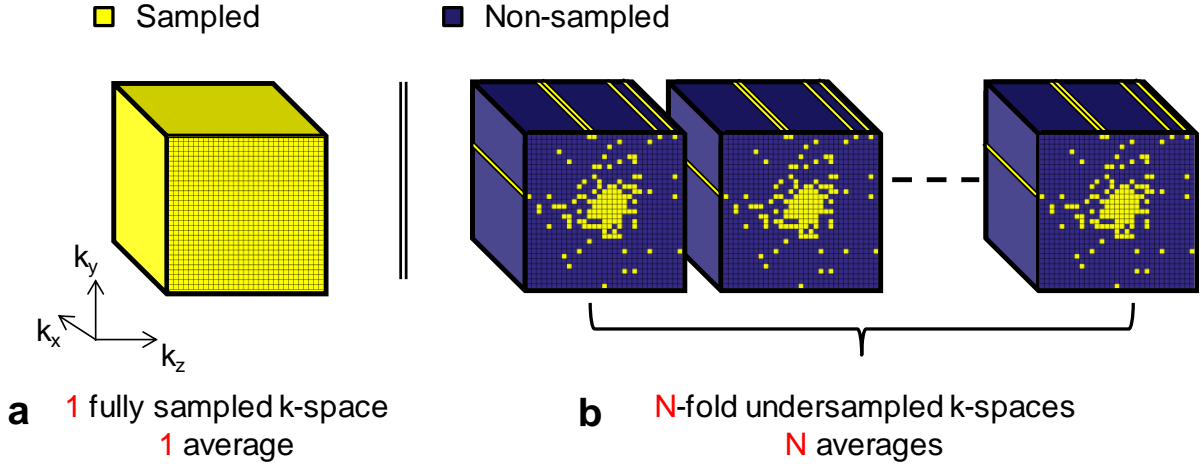

**Figure 1. Schematic overview of the sampling strategies. a.** A fully sampled k-space with each line acquired once. **b.** An N-fold-undersampled k-space, undersampled with a variable density function that fully sampled its center and gradually undersampled its periphery, but with each line acquired N-times. Both sampling strategies **a** and **b** have the same total sample count.

# Figures

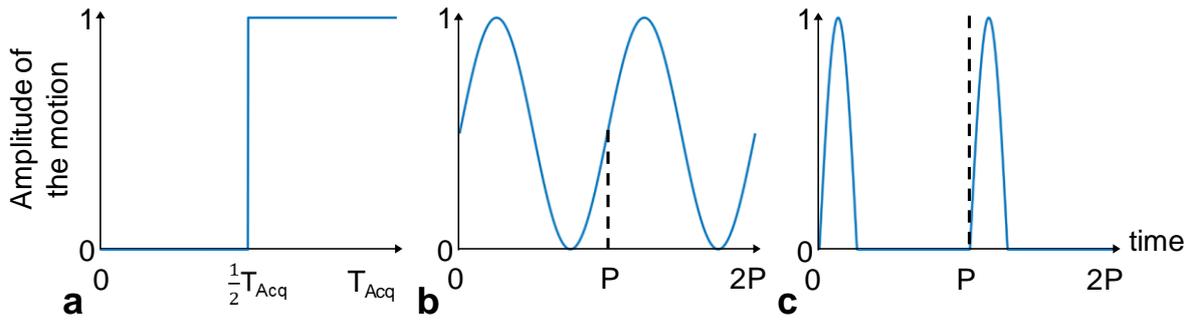

**Figure 2. Motion patterns applied to the phantom.** All patterns were both retrospectively applied to a static dataset via numerical simulation (simulated motion), and prospectively applied via a pump connected to a water reservoir under the phantom (real motion). **a.** A sudden translational motion of the entire subject (body motion). The motion is applied at half the acquisition time ($T_{acq}$). **b**. A periodic sinusoidal motion (sine motion). **c**. Breathing motion: the applied motion models a breathing regime, with a short inspiration (30% of motion) and a longer constant end-expiration.

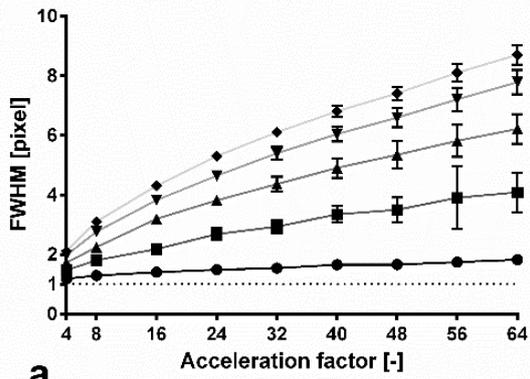
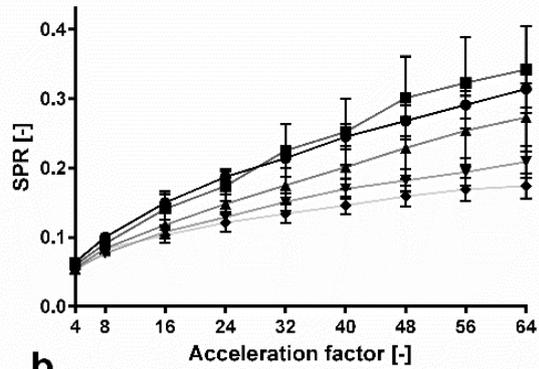
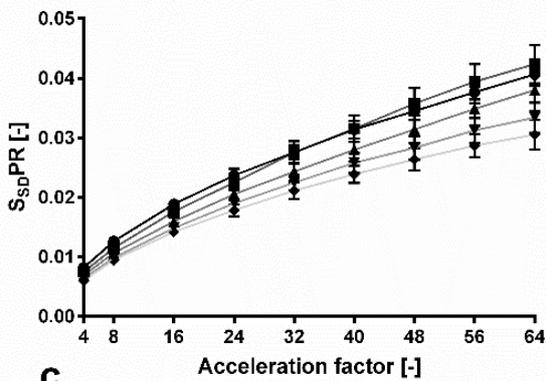
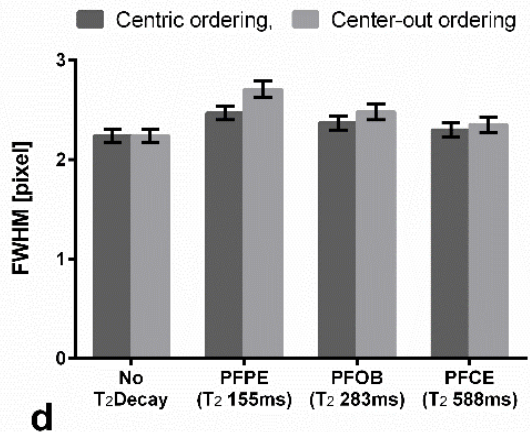

**Figure 3. Characterization of undersampled acquisition patterns through their point spread function. a.** Full width at half maximum (FWHM), **b.** sidelobe-to-peak ratio (SPR) and **c.** sidelobe standard deviation to main peak ratio ($S_{SD}PR$) of the PSF of 45 simulated undersampling patterns without iterative reconstruction. Undersampling patterns were defined through their acceleration factor and FSkC. **d.** FWHM of the PSF of the simulated centric and center-out trajectories with included $T_2$ relaxation of several perfluorocarbons. The coherent effect of the $T_2$ relaxation on the center-out trajectory results in a slightly higher FWHM than for the centric trajectory.

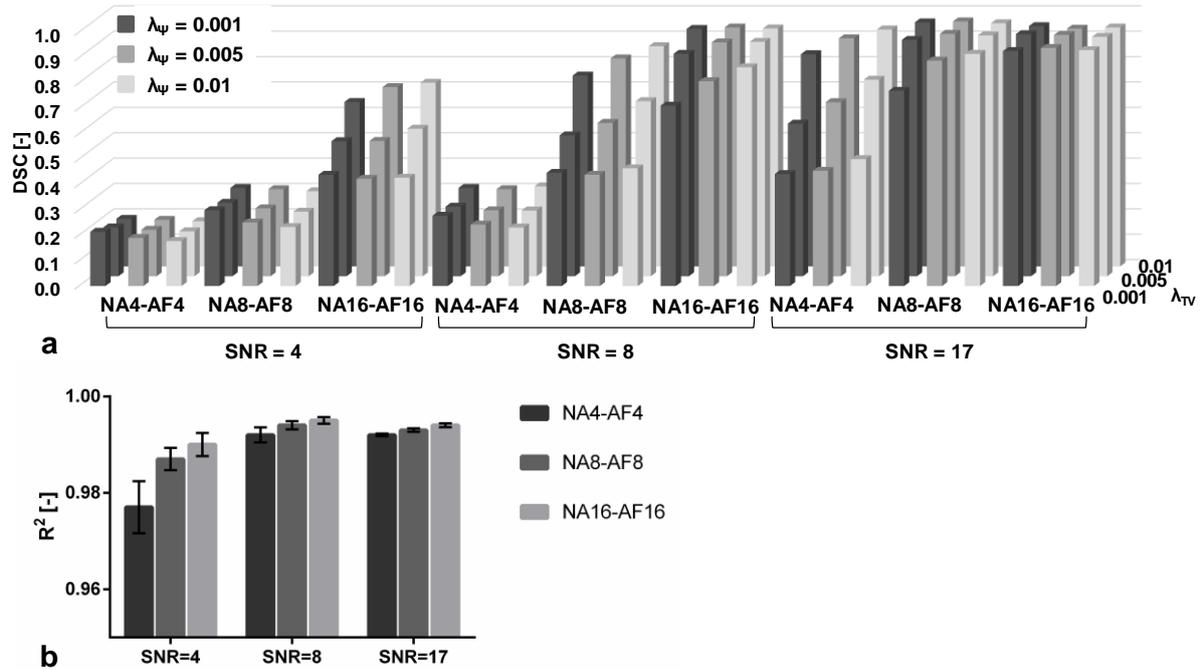

**Figure 4. Dice similarity coefficient (DSC) and coefficient of determination ($R^2$) of the different combinations of reconstruction parameters at three different acceleration factors (AF) and several noise levels. a.** The DSC was calculated for undersampling patterns with a signal-to-noise ratio (SNR) 17, SNR=8, and SNR=4. All undersampling patterns were created with FSkC 25%. For the reconstruction, 9 combinations of the regularization parameters $\lambda_{TV}$ and $\lambda_\psi$ were used, while $\lambda_{ID}$ was fixed at 0.01. **b.** $R^2$ was calculated from the fit of the signal intensity of each phantom tube as a function of the PFPE concentration. At all tested SNR levels, the DSC strongly depended on the regularization parameters and acceleration factors. Except at SNR=17, acceleration factor 16 provided the highest DSCs and $R^2$ of all reconstruction parameters combinations. The reference image was the NA=64 acquisition that was denoised through a wavelet denoising filter with $\lambda_{TV}$=0.05 and $\lambda_\psi$=0.05.

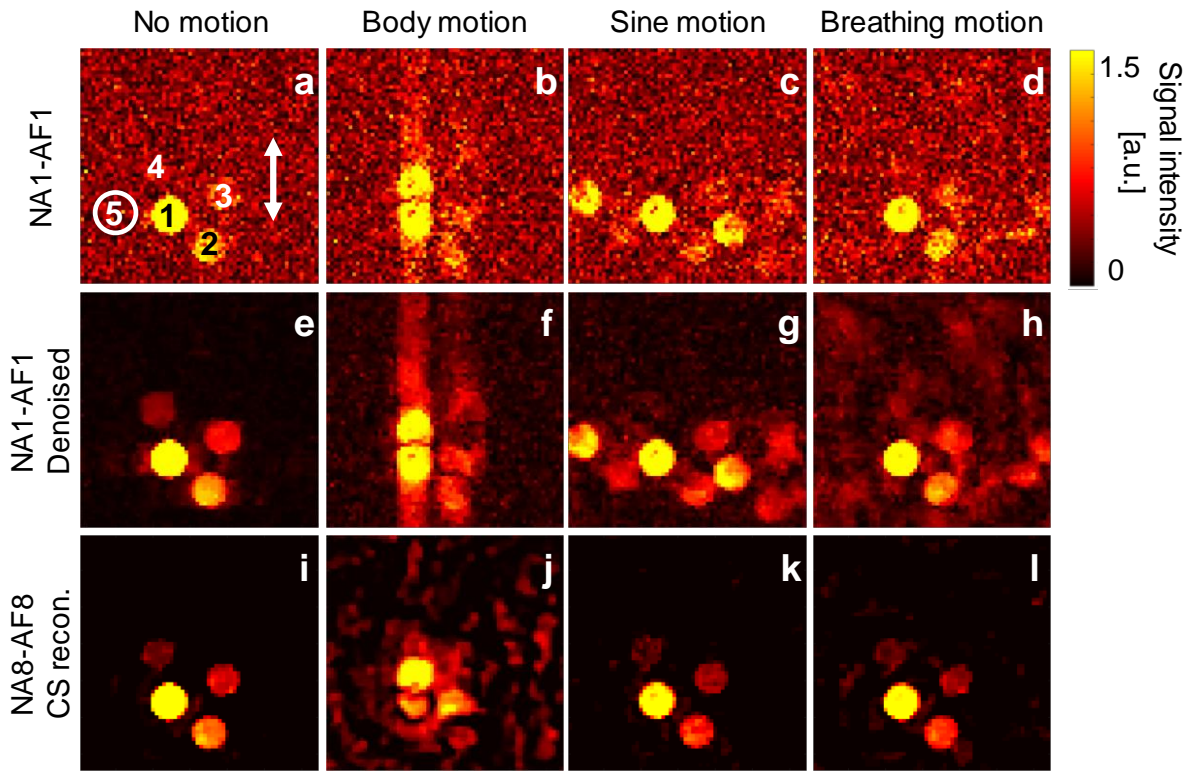

**Figure 5. Phantom images after application of different simulated motion patterns.** Three different motion patterns were applied: body motion, sine motion, and breathing motion. **a-d.** A fully sampled non-averaged $^{19}$F MR acquisition (NA1-AF1) without denoising and **e-h**. with denoising. **i-l**. An 8-fold undersampled $^{19}$F MR acquisition, 8 times averaged (NA8-AF8). The white arrow indicates the direction of the motion. The NA8-AF8 strategy had better robustness against motion than the NA1-AF1-denoised strategy when cyclic motion patterns were applied: only a small amount of remaining background noise can be observed in the NA8-AF8 image compared to its reference, while ghosting artifacts that were not distinguishable from the real phantom signal with sine motion for instance were visible in the denoised strategy.

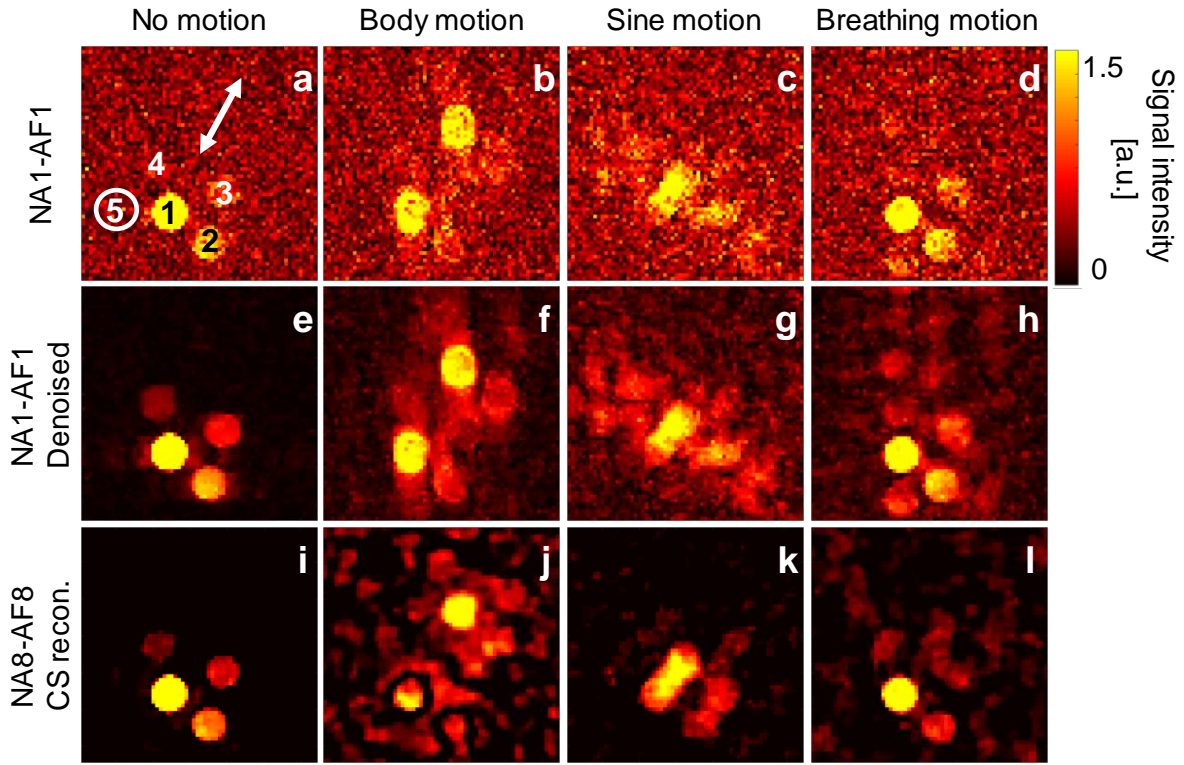

**Figure 6. Phantom images after application of different real motion patterns.** Three different motion patterns were applied: body motion, sine motion and breathing motion. **a-d.** A fully sampled non-averaged $^{19}$F MR acquisition (NA1-AF1) without denoising and **e-h**. with denoising. **i-l**. An 8-fold undersampled $^{19}$F MR acquisition, 8 times averaged (AF8-NA8). The white arrow indicates the direction of the motion, which is diagonal due to the pumping mechanism. Both the denoised and CS-reconstructed images have less background signal than the baseline images. Various types of motion artifacts can be observed in all images acquired during motion, although they differ in size and coherence between the three strategies.

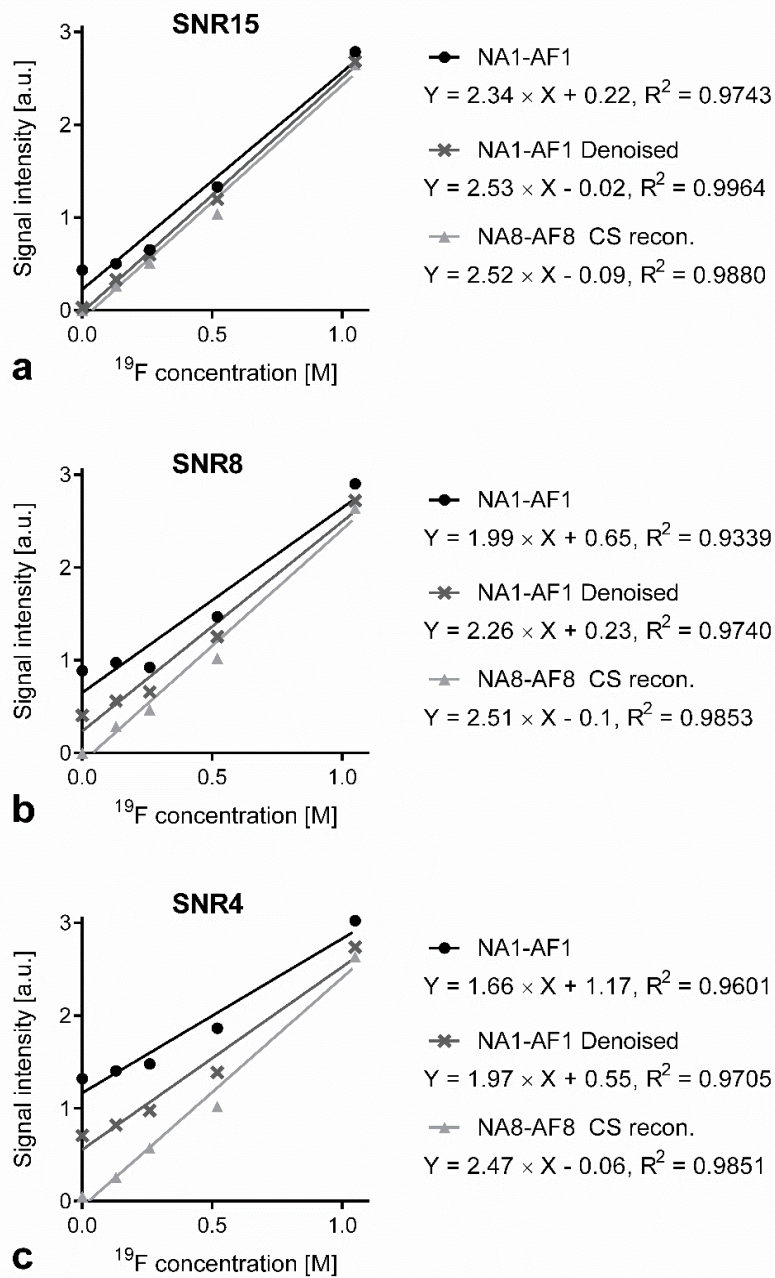

Figure 7. Linear fits of the measured signal intensity versus $^{19}$F concentration for the three strategies: NA1-AF1, NA1-AF1 denoised and NA8-AF8 with CS reconstruction. Linear fits calculated at a. SNR 15, b. SNR 8, c. SNR 4.

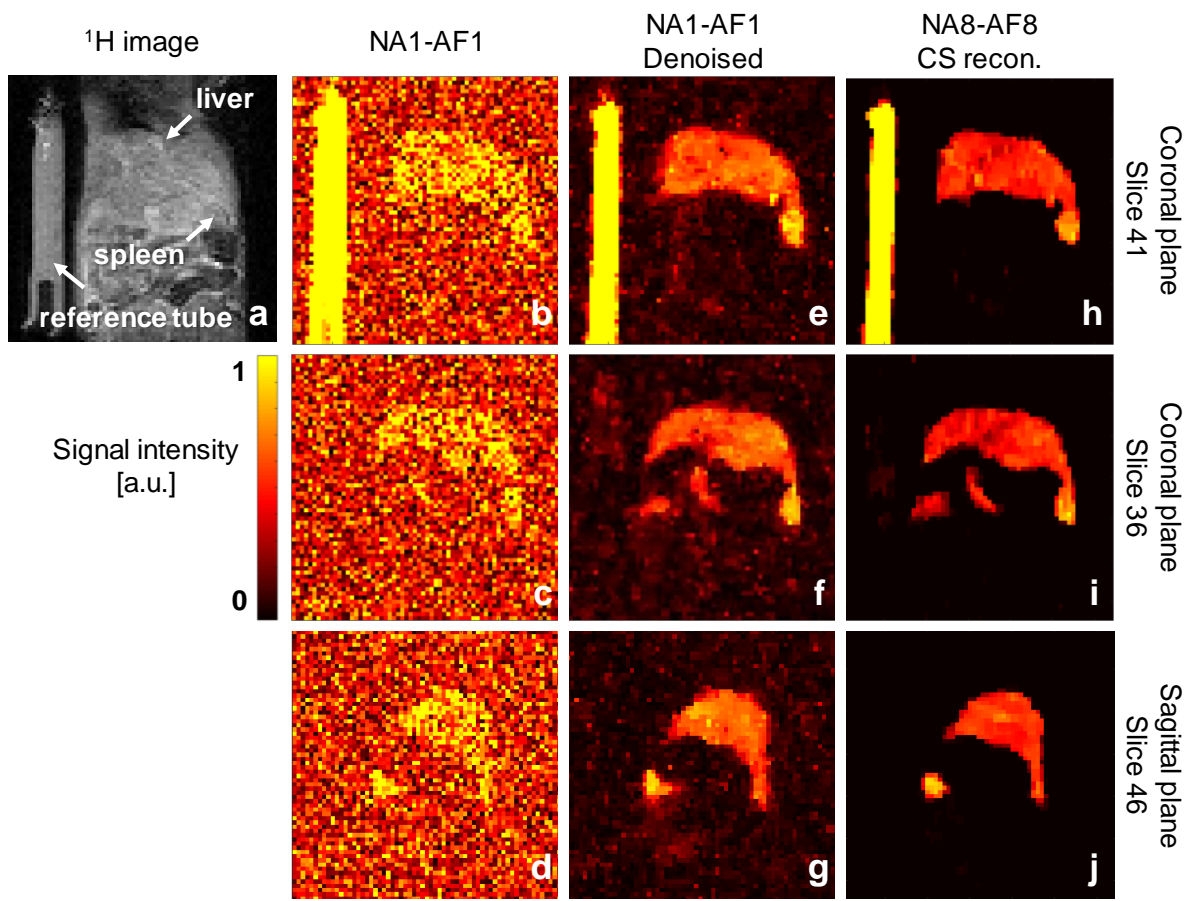

**Figure 8. In vivo images of the mouse abdomen with three strategies. a.** Coronal $^1$H gradient echo image of the mouse abdomen. **b-c-d.** a NA1-AF1 $^{19}$F image without denoising and **e-f-g.** with denoising reconstruction. The liver and spleen of the mouse are more clearly visible in the latter. **h-i-j.** NA8-AF8 $^{19}$F image with CS reconstruction; the liver and spleen are more detailed and there is less residual background signal than in their denoised counterparts.

# Supporting Information

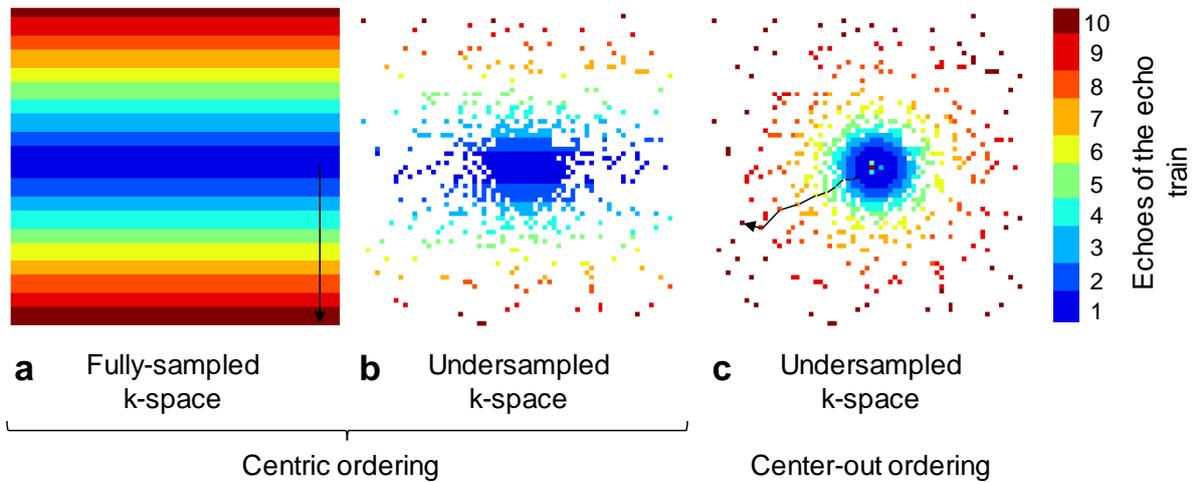

**Supporting Information Figure S1. Illustration of the used centric and center-out k-space trajectories. a.** A fully sampled k-space and **b.** an undersampled k-space with a centric trajectory **c.** An undersampled k-space with a center-out trajectory. The black arrows indicate the direction in which subsequent echoes of the echo train are sampled. Both undersampled k-spaces were obtained with an acceleration factor of 8 and a FSkC of 25%.

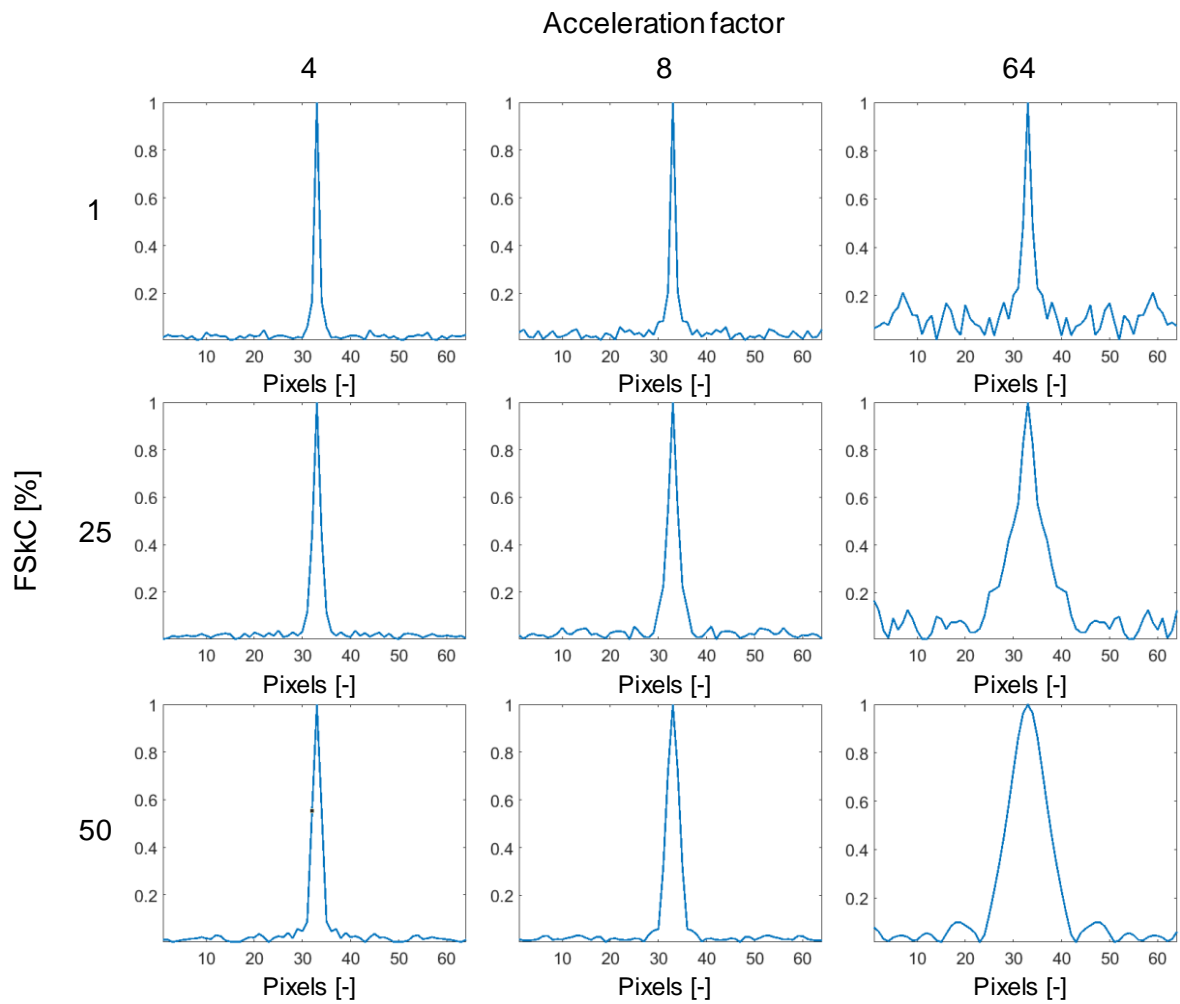

**Supporting Information Figure S2. 1D view of the PSF with varying acceleration factors and FSkC.** While the influence of the FSkC parameter is limited, the higher the acceleration factor, the wider the main peak of the PSF.

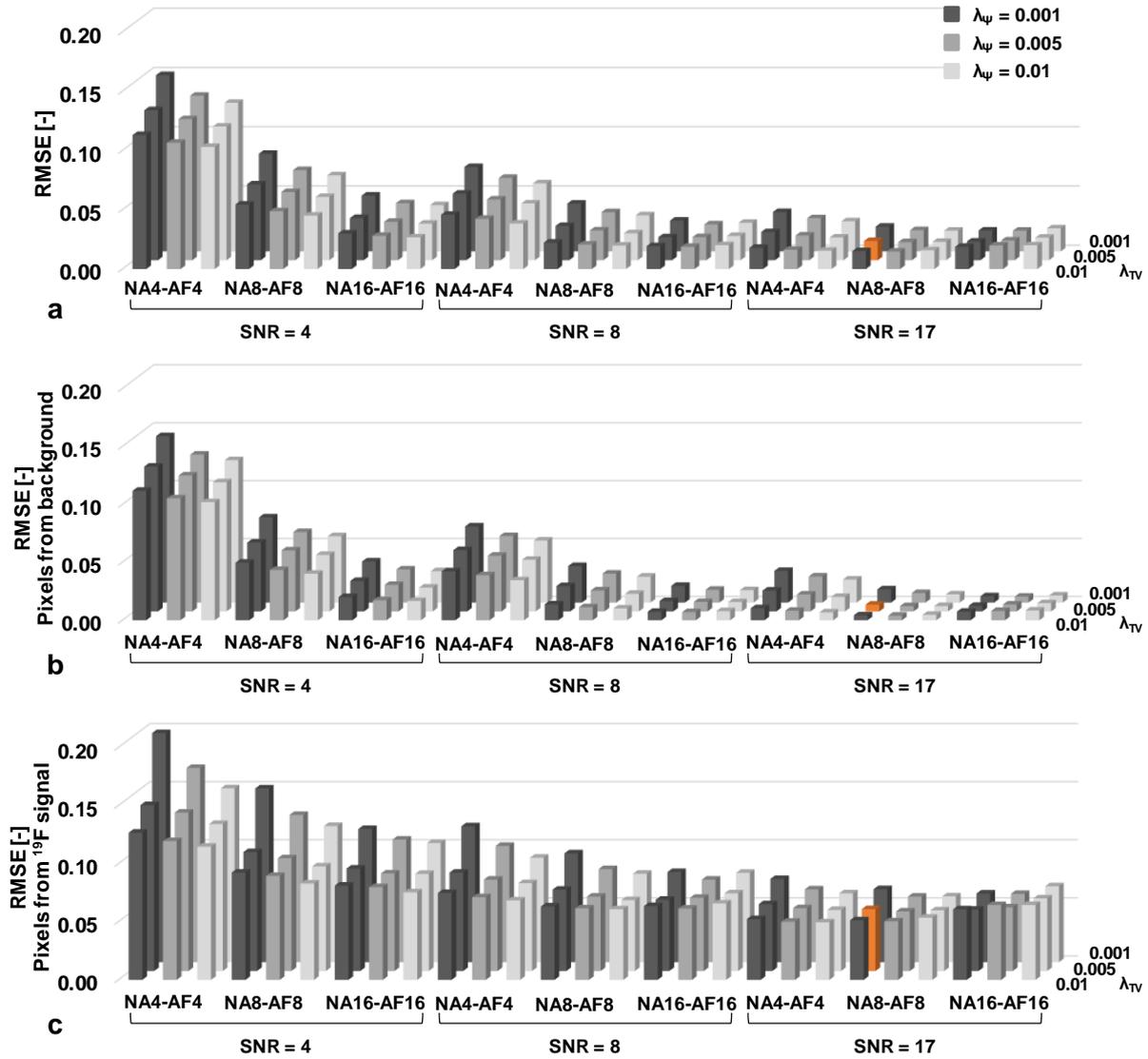

**Supporting Information Figure S3. Root mean square error (RMSE) of the different combinations of reconstruction parameters at three different acceleration factors (AF) with corresponding averages and several noise levels.** The RMSE was calculated for undersampling patterns with a signal-to-noise ratio (SNR) 17, SNR=8, and SNR=4. All undersampling patterns were created with FSkC 25%. For the reconstruction, 9 combinations of the regularization parameters $\lambda_{TV}$ and $\lambda_\psi$ were used, while $\lambda_{ID}$ was fixed at 0.01. At all tested SNR levels, the RMSE strongly depended on the regularization parameters and acceleration factors. A lower value indicates higher image quality. The behavior of the different acquisition-reconstruction values confirms the DSC results. The RMSE was calculated for **a.** the entire image, **b.** the pixels in the background of the image and **c.** pixels of the $^{19}$F signal. As expected, the stronger the regularization terms and the higher the SNR, the lower the RMSE of the background

pixels. However, while the $^{19}$F signals RMSE decreases with the increase in regularization, it remains much higher than that of the background. The orange bar indicates the $\lambda_{TV}$ and $\lambda_{\psi}$ parameters chosen for the in vivo reconstruction – the higher RMSE in the $^{19}$F signal pixels is clearly visible.

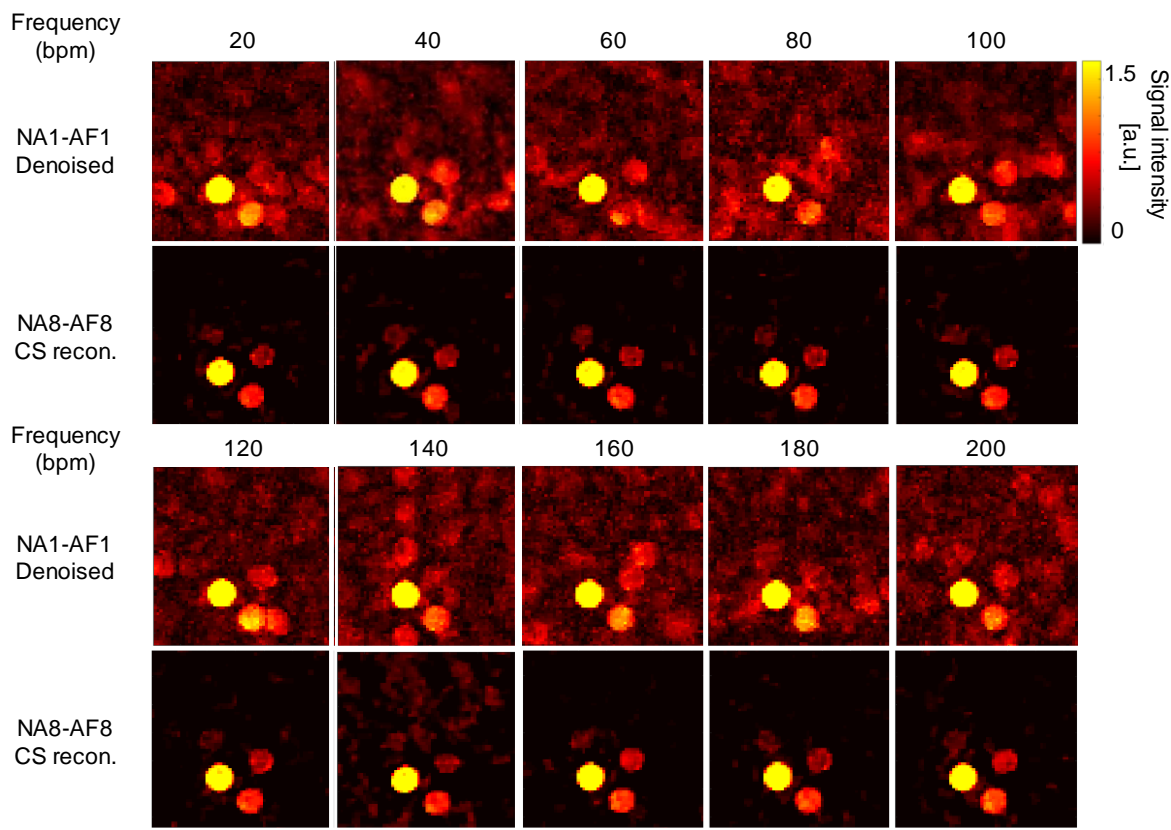

**Supporting Information Figure S4. The influence of the motion frequency on the simulated breathing motion.** Breathing motion was retrospectively applied to NA1-AF1 denoised and NA8-A8 strategies with a fixed amplitude of 30% of the FOV and a frequency that varied from 20bpm (breaths per minute) to 200 bpm. While ghosting tubes are visible in the NA1-AF1 images, they are not visible in the NA8-AF8 strategy, except at 140 bpm.

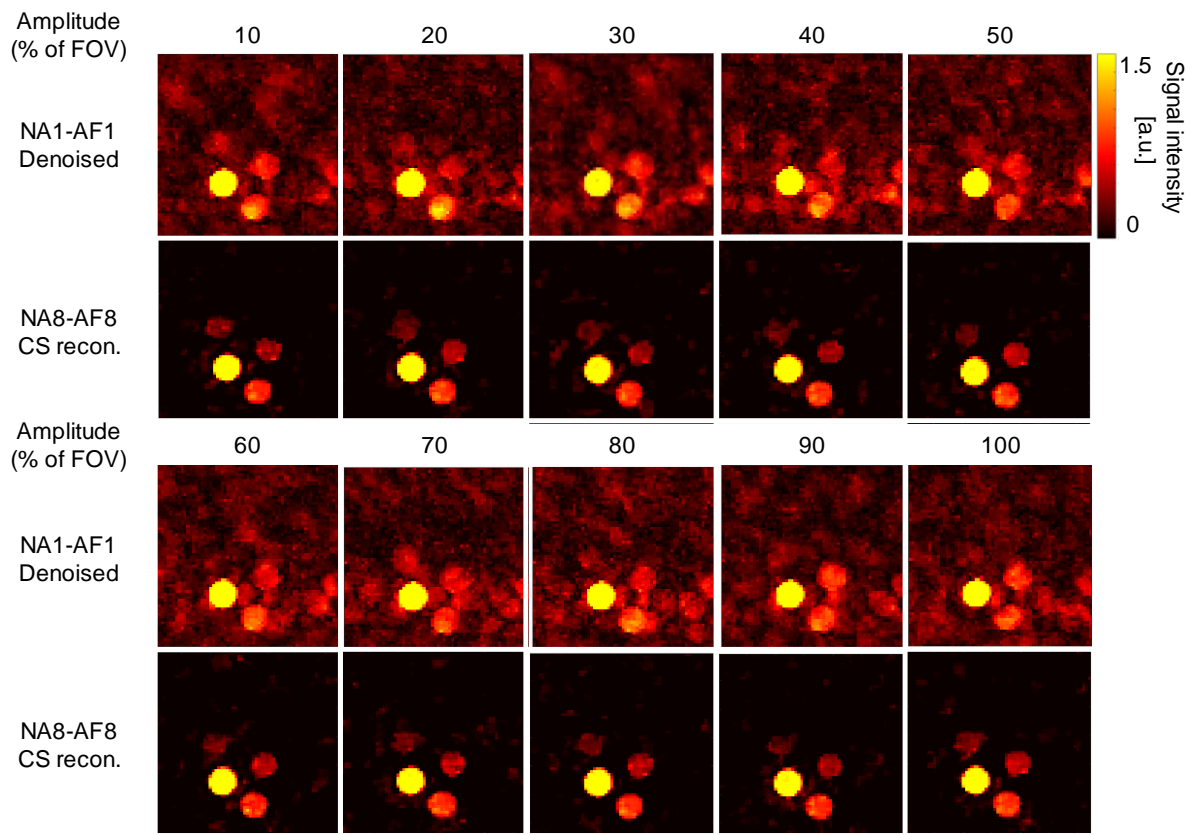

**Supporting Information Figure S5. The influence of the motion amplitude on the simulated breathing motion.** Breathing motion was retrospectively applied to NA1-AF1 denoised and NA8-A8 strategies with a fixed frequency of 40 bpm and an amplitude that varied from 10 to 100 % of the FOV. The amplitude variation does not affect the images of either sampling strategies.

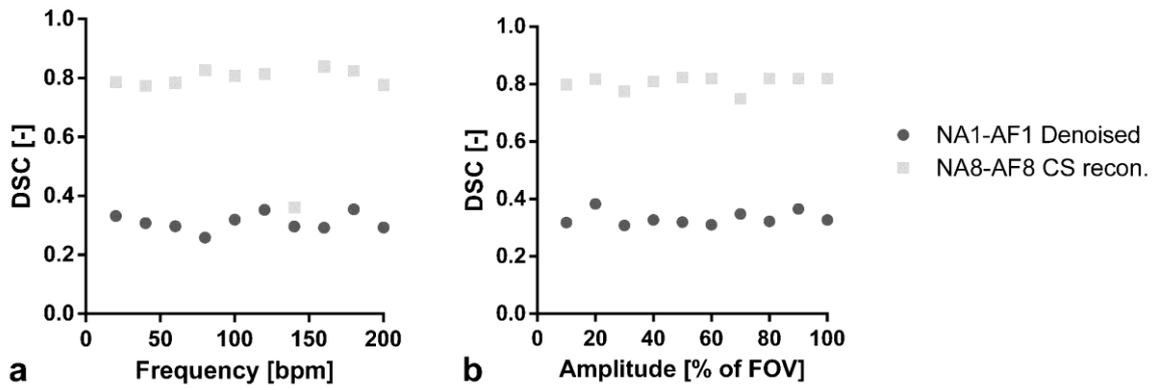

**Supporting Information Figure S6. Dice similarity coefficients between the moving images and their respective static counterparts for a range of motion frequencies and motion amplitudes. a.** DSC of NA1-AF1 and NA8-AF8 with the motion amplitude fixed at 30% of the FOV and **b.** with the motion frequency fixed at 40 bpm.

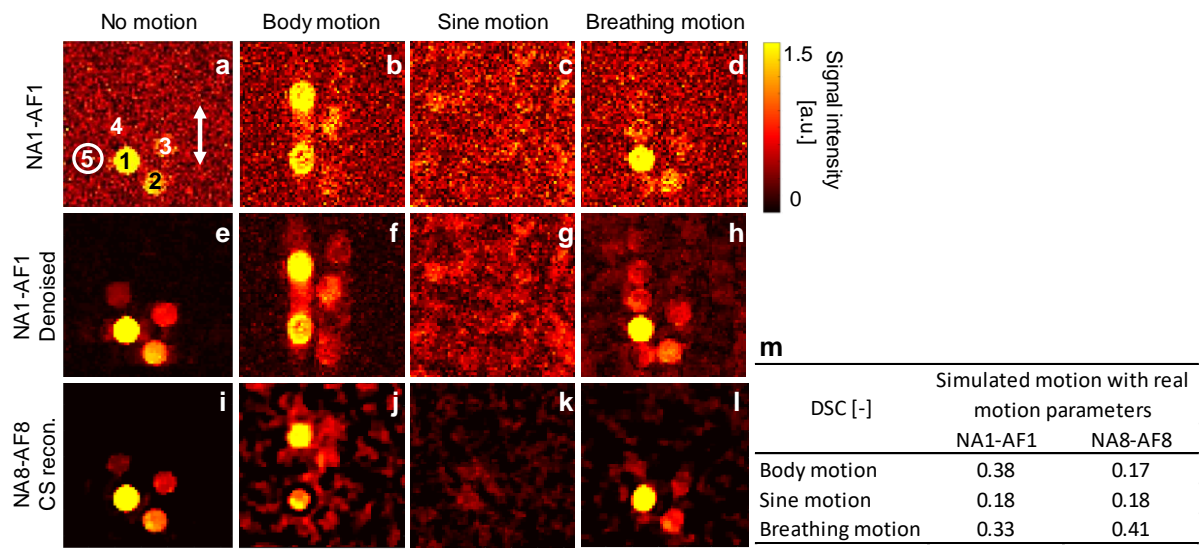

**Supporting Information Figure S7. Phantom images after application of simulated motion patterns with the frequencies and amplitudes used in the real motion experiment.** Three different motion patterns were applied: body motion, sine motion, and breathing motion. **a-d**. A fully sampled non-averaged $^{19}$F MR acquisition (NA1-AF1) without denoising and **e-h**. with denoising. **i-l**. An 8-fold undersampled $^{19}$F MR acquisition, 8 times averaged (NA8-AF8). The white arrow indicates the direction of the motion. The NA8-AF8 strategy had better robustness against motion than the NA1-AF1-denoised strategy when breathing motion pattern was applied: only a small amount of remaining background noise can be observed in the NA8-AF8 image compared to its reference. With the sine motion, no signal can be distinguished in any of the images. **m.** The DSCs were calculated between each image with induced motion and their corresponding static image, where both were reconstructed with the same reconstruction parameters. DSCs of body and breathing motion were significantly different between the two sampling strategies (P<0.001), DSC of Sine motion were not (P = 0.3).